\documentclass[10pt]{article}
	\usepackage{latexsym,graphicx,multirow}
	\usepackage{amssymb}
	\usepackage{amsmath}
	\usepackage{amscd}
	\usepackage{amsthm}
	\usepackage[left=2cm,top=2.5cm,right=2.5cm,bottom=1.5cm]{geometry}
	\usepackage{hyperref}
	\usepackage{epstopdf}
	\usepackage{float}
	\usepackage{subcaption}
	
\begin{document}

	\begin{center}
	\large{\bf{Compatibility between the scalar field models of tachyon, $k$-essence and quintessence in $f(R, T)$ gravity }} \\
	\vspace{10mm}
	\normalsize{ Vinod Kumar Bhardwaj$^1$, Anirudh Pradhan$^2$, Archana Dixit$^3$  }\\
	\vspace{5mm}
	\normalsize{$^{1,2,3}$Department of Mathematics, Institute of Applied Sciences and Humanities, G L A University\\
		Mathura-281 406, Uttar Pradesh, India}\\
	\vspace{2mm}
	$^1$E-mail: dr.vinodbhardwaj@gmail.com\\
	$^2$E-mail: pradhan.anirudh@gmail.com\\
	$^3$E-mail: archana.dixit@gla.ac.in\\
	
	\vspace{10mm}
	
	%\date{}
	%\maketitle
	\end{center}
        \begin{abstract}
	
	We investigate the behavior of the scalar field in $f (R, T )$ gravity (Harko et al., Phys. Rev. D 84, 024020, 2011) 
	inside the structure of a flat FRW cosmological model, where $R$ and $T$ have their usual meaning. The deterministic solution of 
	FEs  has been settled by considering the scale factor $S(t) = t^{m}e^{kt}$, where $m$ and $k$ are positive constant. Here we utilize three 
	ongoing imperatives ($H_{0}=73.8$ and $ q_{0}=-0.73$) from SNe Ia Union observation (Cunha, Phys. Rev. D 79, 047301, 2009); 
	($H_{0}=73.8$ and $ q_{0}=-0.55$) from Type Ia Supernova (SNIa) observation in joined with (BAO) and (CMB) observation 
	(Giostri et al., J. Cosmol. Astropart. Phys. 03, 027, 2012); and ($H_{0}=69.2$ and $q_{0}=-0.52)$ from (OHD+JLA) observation 
	(Yu et al., Astrophys. J. 856, 3, 2018). We have measured and graphed several cosmological parameters and found that these 
	findings are appropriate in comparison with the universe's physical and kinematic properties and also compatible with ongoing 
	observations. The correspondence between the scalar field models of tachyon, k-essence, and quintessence is investigated by us. 
	Our model explains the potentials and the dynamics of the scalar fields in the FRW universe. The reconstructed potential  is
	very sensible and has a scaling arrangement. Subsequently, our study supports recent observations.
	
	\end{abstract}
	

	%%%%%%%%%%%%%%%%%%%%%%%%%%%%%%%%%%%%%%%%%%%%%%%%%%%%%%%%%%%% Section - Introduction %%%%%%%%%%%%%%%%%%%%%%%%%%%%%%%%%%%%%%%%%
	
\section{Introduction}

The Scientific cosmology is intended to determine the  broad-scale structure of the universe. The cosmic perceptions of type-Ia supernovae 
tests \cite{ref1}-\cite{ref7} propose that the observable universe is going through rapid growth. The ongoing observational information 
unequivocally shows that the models of the universe, obtained in this paper, are prevailed by a segment with negative pressure, named as 
dark energy (DE), which comprises three fourth of the basic critical density. In order to explain the concept of dark energy and accelerated 
expansion, the decisions of certain hypothetical models must be unique to quintessence scalar field models \cite{ref8}-\cite{ref9}, 
k-essence \cite{ref10}-\cite{ref11}, tachyon field \cite{ref12}-\cite{ref13}, phantom field \cite{ref14}-\cite{ref15}, 
quintom \cite{ref16}-\cite{ref17} as well as dark energy models like chaplygin gas \cite{ref18}-\cite{ref19}. Many cosmologists were studying 
cosmological theories of dark energy in alternate gravitational speculations.\\

Regarding their cosmological implications, altered theories of gravitation have been broadly concentrated. Numerous authors 
\cite{ref20}-\cite{ref23} were studying cosmological theories of dark energy in modified gravitational speculations.. In recent years 
there has been a growing interest in modified gravity theories despite the clear evidence of  the late-time acceleration of the Universe  and 
the presence of dark matter and energy.  \\

Several ideas had been developed and reviewed in the literature. Among these, F(R) gravity theory is an appropriate candidate theory 
that analyzed Einstein's (GR) theory by supplanting the gravitational action $R$ with the arbitrary function $f(R)$. This theory gives 
more perplexing conditions than general relativity and gives a larger range of arrangements. Utilizing this theory, Starobinsky \cite{ref24} 
initially proposed an accelerated inflation plan. Sotiriou \cite{ref25} addressed the conditions for the f(R) and scalar-tensor theories to 
be comparable.\\

Various modified theories are discussed in \cite{ref26} their structures have been  examined by analyzing Big Rip and future singularities. 
Huang \cite{ref27} explored an inflation model of $f(R)$. Likewise several cosmologists have also researched $f(T)$ gravity theory in different 
ways. Modified gravity models additionally include $f(T)$ gravity \cite{ref28,ref29}, $f(R)$ gravity \cite{ref30,ref31}, Gauss-Bonnet 
gravity \cite{ref32}, scalar-tensor speculations \cite{ref33,ref34}, Galileon gravity \cite{ref35}, Braneworld models \cite{ref36} and so on.\\

In the same context, several authors have extensively researched the astrophysical and cosmological implications of $f(R, T)$ gravity. 
Many researchers are recently involved in confronting cosmological string models in the adjusted scenario for $f(R, T)$ gravity theories 
\cite{ref37,ref38}. Authors \cite{ref39,ref40,ref41} have studied the physical aspects of cosmological models in updated $f(R, T)$ 
gravity in various contexts. Sharif et al. \cite {ref42} examined  the energy conditions for FRW Universe with perfect fluid in $f(R, T)$ 
gravity. Houndjo \cite{ref43} has arranged a cosmological rehabilitation of $f(R, T)$ gravity which handles the transition from a 
matter-dominated phase to an accelerated phase. Ahmed and Pradhan \cite{ref44} discussed the accelerating cosmological model of 
Bianchi type-V, with $\Lambda(T)$ gravity. In the modified $f(R, T)$ theory of gravity, Singh and Kumar \cite{ref45} have shown the 
effects of bulk viscosity and explained the early and late acceleration of the Universe. Sahoo et al. \cite{ref46} addressed the Kaluza 
Klein cosmological model in $f(R, T)$ gravity where $\Lambda$ is a stress-energy tensor depending on $T$. The future advancement of 
the dark energy universe was explored by Bamba et al. \cite{ref47} in modified gravities incorporating a perfect fluid with the 
inhomogeneous equation of the state in $f(R)$ gravity. There are a large number of works which are impossible to be  mentioned  here but we can 
refer to some latest references in $f(R, T)$ gravity \cite{ref48}-\cite{ref65} (and references therein). \\

On the other hand, the tachyon field has been suggested as a possible candidate for dark energy. A rolling tachyon has an interesting 
equation of state whose parameter smoothly/swimmingly estimates the values between -$1$ and $0$ \cite{ref66,ref67}. Thus, tachyon can be 
realized as a suitable candidate for inflation at high energy \cite{ref68}. This is also treated as a source of DE depending on the 
form of tachyon potential \cite{ref69}. The tachyon has been intensively studied in the last few years also in application to cosmology 
\cite{ref70}-\cite{ref74}. Sheykhi et al. \cite{ref75} has reconstructed the potential and the dynamics of the tachyon field according 
to the evolutionary nature of ghost dark energy. Setare et al. \cite{ref76} investigated the spherical collapse and the evolution 
of over-densities in the context of the tachyon scalar field model. In this article, the authors have also compared the results with the results 
of Einstein-de Sitter and $\Lambda$-cold dark matter models. Recently, Dimitrijevic et al. \cite{ref77} has discussed a class of tachyon models 
in a braneworld cosmology scenario based on the second Randall-Sundrum model extended to more general warp factors. A model of tachyon inflation 
in the framework of holographic cosmology has been recently studied in \cite{ref78}. \\

Another possibility to explain the mystery of dark energy is $k$-essence. It is to work with the theme that the unknown DE constituent 
is due to minimally coupled scalar field $\phi$ with non-canonical kinetic energy \cite{ref79}. In the present time, $k$-essence scenario 
has received much attention. Originally, it was purported as a model of inflation \cite{ref80} and 
then a model of DE \cite{ref79}. Many authors \cite{ref81}-\cite{ref88} have studied DE cosmological models in the framework of $k$-essence 
in various contexts. Sebastiani et al. \cite{ref89} has focused on $f(T)$ gravity and $k$-essence cosmology. Recently, Srivastava et al. \cite{ref90} 
has discussed Bianchi-III universe with $k$-essence for holographic dark energy. Dubey et al. \cite{ref91} has obtained Tsallis HDE 
in the Bianchi-I universe using hybrid expansion law with $k$-essence. \\

Guendelman et al. \cite{ref92} identified a quintessence, unified dark energy and dark matter, and process of containment
where the effective vacuum energy density and dynamically induced dust-like matter appear as dynamically produced, correspondingly.
Aktas¸\cite{ref93} has explored the tachyon, $k$-essence, and quintessence of DE in (FRW) universe models with varying $G$ and $\Lambda$ in 
the theory of $f(R, T)$ gravitation. They used a linearly varying deceleration parameter (LVDP) for the solution of the field equation. 
An advanced cosmological study of $G$ and $\Lambda$ play a significant role because it may be responsible for accelerating the Universe. 
Several authors have thoroughly covered the cosmological constant as a function of time in different variable $G$ theories in various contexts. In 
general relativity (GR), lots of modifications were suggested to allow for a variable $G$ based on various arguments. In a flat FRW universe, 
Norman Cruz\cite{ref94} consider a relation between the holographic dark energy density and the kinetic k-essence energy density.
They defined if $c > 1$, the holographic dark energy is described as a kinetic $k$-essence scalar field that certain way.\\

Granda and Olivere \cite{ref95}  introduced and worked on the relationship between quintessence, tachyon, k-essence, and dilation energy 
density with HDE in the flat FRW universe, adding a new infrared cut-off for the HDE. This relationship allows the models of the scalar 
fields to recreate the potentials and dynamics, which characterize accelerated expansion. The work in \cite{ref96} showed the reconstruction 
of scalar-field energy models for general Lagrangian density $p(\phi,X)$, where  $\phi$ and $X$ is the scalar field and kinematic term.
Sharif and Jawad \cite{ref97,ref98} worked on the reconstruction of scalar field dark energy models in the flat Kaluza-Klein universe. They 
are also developing their correspondence with certain dark energy simulations of scalar fields. One of the most fascinating and challenging 
problems in cosmology is the discovery of the accelerated expansion of the Universe \cite{ref99}-\cite{ref105}. There are many ways to resolve 
this problem, including: the cosmological constant, the quintessence field \cite{ref106}-\cite{ref108}, a brane cosmology scenarios 
\cite{ref109}-\cite{ref110} and models of k-essence \cite{ref111}-\cite{ref115}. The cosmological constant is the easiest option among the earlier 
proposals which requires a fine-tuning value .\\

Considering the scalar field dark energy models as an effective explanation of the underlying dark energy theory, it is fascinating to 
concentrate on how the scalar field models can be utilized to depict the energy density as powerful speculations. In this work, we discuss 
the cosmological parameters in the framework of the $f(R, T)$ gravity and also discussed the evaluation of the scalar field with the tachyon, 
k-essence, and the quintessence models. The present research is motivated mainly by \cite{ref93}, and has also been influenced by 
\cite{ref95, ref116,ref117,ref118}. The paper is structured as follows: In Sect. $2$, we express the dynamics of the field equations on $f(R, T)$. 
Sect. $3$ described the solution of the field equations. We discussed the correspondence of the scalar field, in Sect. $4$ and this section 
can be divided into three parts tachyon, k-essence, and quintessence fields with redshift. Finally, the results of the models are summarized. 
in Sect. $5$.
	
%%%%%%%%%%%%%%%%%%%%%%%%%%%%%%%%%%%%%%%%%%%%%%%%%%%%%%%%%% Section 2 %%%%%%%%%%%%%%%%%%%%%%%%%%%%%%%%%%%%%%%%%%%%%%%%%%%%%%%%%%%%%%%%%%%
\section{ Dynamics of the Field Equations}

Harko et al. \cite{ref119} proposed the modified theory of gravity in the form of $ f(R,T) $ theory  to describe the accelerated expansion 
of universe. The altered conditions are likewise comprehended for fluctuating $\Lambda$ and $ G $. To give the arrangement of $ f(R,T) $ 
gravity recommended by Harko et al. \cite{ref119} following three $ f(R,T) $ models

\begin{equation}
\label{1}
f(R,T) =\left\{
\begin{split}
&R+2\lambda_{1}(T), \\
&f_{1}(R)+f_{2}(T),\\
&f_{R}+f_{2}(R)f_{3}(T).
\end{split}\right.	
\end{equation}

 The adjusted field conditions in $ f(R,T) $ theory with variable  $ \Lambda $ and $ G $  are given by Tiwari et al. \cite{ref74} as

\begin{equation}
\label{2}
G_{\alpha \beta}=\left[8 \pi G(t)+2\lambda'_{1}(T)\right]T_{\alpha \beta}+\left[2 p \lambda'_{1}(T) +\lambda_{1}(T)+\Lambda(t)\right]g_{\alpha \beta},
\end{equation}

where `dash' implies differentiation.
We select $\lambda_{1}(T)=\mu T $, Eq.(\ref{2}) can be inferred as Tiwari et al. \cite{ref120}, here ( $\mu  \rightarrow $ constant).

\begin{equation}
\label{3}
G_{\alpha \beta}=\left[8 \pi G(t)+2\mu \right]T_{\alpha \beta}+\left[\mu \rho-\mu  p +\Lambda(t)\right]g_{\alpha \beta},
\end{equation}

 In the current study, we will discuss about the different dark energy models in FRW space-time with variable $ \Lambda $ and $ G $ in 
 adjusted $ f(R,T) $ gravity theory. To acquire GRT arrangements, we get $ \mu=0 $ in Eq. (\ref{3}). The flat FRW universe is given by

\begin{equation}
\label{4}
ds^{2}=dt^{2}-S^{2}\left[dr^{2}+r^{2}(d\theta^2+\sin^{2}\theta  \ d\phi^2)\right].
\end{equation}

 the Ricci scalar is defined for  flat FRW universe  as:

\begin{equation}
\label{5}
R=-6\left(\frac{\ddot{S}}{S}+\frac{\dot{S}^2}{S^2}\right)
\end{equation}

 Here we defined energy-momentum tensor as:

\begin{equation}
\label{6}
T_{\alpha\beta}= -p g_{\alpha\beta}+(\rho+p)u_{\alpha}u_{\beta}
\end{equation}

where $ \rho $ and $ p $ represents the energy density and pressure respectively. $ u^{i}=(0,0,0,1) $ indicates the four-velocity vector 
components. $ T^{DF}=\rho-3p $ is the trace of energy-momentum tensor \\

The field equations of a dark energy model for flat FRW universe in $ f(R,T)$ theory of  gravity are expressed as:

\begin{equation}
\label{7}
3 H^2+2 \dot{H}=-8\pi G p +\mu \rho-3\mu p+\Lambda
\end{equation} 

\begin{equation}
\label{8}
3 H^2=8\pi G \rho-\mu p+3\mu \rho+\Lambda
\end{equation}  

Here, $ H=\frac{\dot{S}}{S} $ is the Hubble parameter.
%%%%%%%%%%%%%%%%%%%%%%%%%%%%%%%%%%%%%%%%%%%%%%%%%%%%%%%%%%%%%%%%%%% Section 3 %%%%%%%%%%%%%%%%%%%%%%%%%%%%%%%%%%%%%%%%%%%%%%%%%
\section{ Solution of the Field Equations}
 We have two field  equations as mentioned by (\ref{7}) and (\ref{8}), which involves five unknowns as $H, \rho$,  $p$, $G$, $\Lambda$.  
%To get the solution of field equations, we consider the following assumptions
Thus, for deterministic solution, we need three more assumptions as
\begin{itemize}
\item[(i)] we take the ratio between $ H^2 $ and $ \Lambda $, i.e.$ \Lambda=\xi H^2 $,  Schutzhold \cite{ref121} here $ \xi $ is a constant. 
The cosmological term is a variable that depends on the values of $G, \rho, \ H^2 $and $\dot{H}$. 
These equations are related to $H$ for a known value of $\Lambda$ and many authors have been worked on this assumption \cite{ref122,ref123} .
\item[(ii)] Expression of EoS parameter between energy density and the pressure as $ \omega=\frac{p}{\rho} $.
\item[(iii)] We assume a deceleration parameter as:
\begin{equation}
\label{9}
q=-\left(\frac{\dot{H}+H^2}{{H}^2}\right)=\frac{m }{(m+kt)^2}-1
\end{equation} 
\end{itemize}

From Eq. (\ref{9}), we observed that $ q>0 $ for $ m(1-m)>2m k t+k^2 t^2 $ and $ q=-1 $ for $ m=0 $. It is observed that for 
$ k=0 $, $ q=\frac{1}{m}-1 $, in this case $ q>0 $ or $ q<0 $ according as $ m<1 $ or $ m>1 $ respectively.

On solving Eq. (\ref{9}), we get metric potential as 
\begin{equation}
\label{10}
S=t^m exp(kt)
\end{equation}  
where $ m $ and $ k $ are constants. 

Eq. (\ref{9}) defines a relationship between constants $k$ and $m$ for the present universe ($ t_{0}=13.8 $Gyr) with 
$q_{0} = -0.73 $ (Cunha \cite{ref124}).
\begin{equation}
\label{11}
k=\frac{1}{13.8}\left[\sqrt{\frac{m}{0.27}}-m\right]
\end{equation}
  
From Eq. (\ref{11}), it is obvious that model for the present universe is true for $m > 0.27$. It is observed that our model is in 
the accelerating phase for $m > 1,$ and model is in the transition phase for $ 0.27 < m \leq 1 $. \\

From Eq. (\ref{9}), we obtain a relation between $k$ and $m$ for the present universe ($ t_{0}=13.8 $ Gyr) with $ q_{0}=-0.55 $ 
(Giostri et al. \cite{ref125}), we again a relation as:

\begin{equation}
\label{12}
k=\frac{1}{13.8}\left[\sqrt{\frac{m}{0.45}}-m\right]
\end{equation}

Eq.(\ref{12})explains that the model  is appropriate for $m \geq 0.45 $. It is also clear from the relation that our model is in 
the transition phase for $ 0.45 < m \leq 1 $ and the accelerating phase for $m > 1,$.

Similarly, from Eq. (\ref{9}), we find a relation between $k$ and $m$ for the present universe $ t_{0}=12.36 $ and $ q_{0}=-0.52 $
(Amirhaschi \& Amirhaschi \cite{ref126}); Yu et al. \cite{ref127}) as: 

\begin{equation}
\label{13}
k=\frac{1}{12.36}\left[\sqrt{\frac{m}{0.48}}-m\right]
\end{equation}

The relation expresses that the model is valid for $m \geq 0.48 $ and also represent the accelerated expansion of the universe for $ m > 1 $. \\

The various estimations of pair ($k$, $m$) are obtained from Eqs. (\ref{11})-(\ref{13}) as per three observational imperatives 
\cite{ref124,ref125,ref126,ref127} and are given in Table-1. These experimental estimations of $ k $ and $ m $ 
are utilized for plotting and validation of the derived models.
%%%%%%%%%%%%%%%%%%%%%%%%%%%%%%%%%%%%%%%%%%%%%%%%%%%%%%%%%% Table 1 %%%%%%%%%%%%%%%%%%%%%%%%%%%%%%%%%%%%%%%%%%%%%%%%%%%%%%%%
\begin{table}[htb] 
	\centering
	\begin{tabular}{|c|c|c|c|}
		%\hline\hline %inserts double horizontal lines
		\hline
		$ m $ & $ k $ (Cunha et al. 2009) & $ k $ (Giostri et al. 2012) & $ k $ (Yu et al. 2018)  \\ [1ex]
		\hline
		0.3 & 0.0546443879 & 0.03742728847 & 0.03969008212  \\
		\hline
		0.5 & 0.06237881413 & 0.04015163428 & 0.04212141796 \\
		\hline
		0.8 & 0.06676274870 & 0.03864734297 & 0.0397244699  \\
		\hline
		1.0 & 0.06699281858 & 0.03555883948 & 0.03587181823  \\
		\hline
		1.5 & 0.0621030872 & 0.0236044824 & 0.0216639930  \\
		\hline
		2.0 & 0.0522938602 & 0.0078395005 & 0.0033366871  \\
		\hline
	\end{tabular}
	\caption{ Table of Values of $ m $ and $ k $} % title of table
	\label{table1} %is used to refer this table in the text
\end{table}
%%%%%%%%%%%%%%%%%%%%%%%%%%%%%%%%%%%%%%%%%%%%%%%%%%%%%%%%%%%%%%%%%%%%%%%%%%%%%%%%%%%%%%%%%%%%%%%%%%%%%%%%%%%%%%%%%%%%%%%%%%%%%%%%%%%%%%%%%

By using Eq. (\ref{7}), Eq. (\ref{8}) and Eq. (\ref{10}), EoS parameter and $\Lambda=\xi H^2, $ for this model, the dynamics of the 
universe are determined as:

\begin{equation}
\label{14}
\rho=\frac{(\xi-3)(1+\omega)(m+k t)^2+2 m}{\mu (\omega^2-1) t^2}
\end{equation}  
\begin{equation}
\label{15}
p=\frac{\omega\left[(\xi-3)(1+\omega)(m+k t)^2+2 m\right]}{\mu (\omega^2-1) t^2}
\end{equation} 

\begin{equation}
\label{16}
G=\frac{\mu\left[(3-\xi)(1+\omega)(m+k t)^2+2 m (\omega-3)\right]}{4\pi\left[(\xi-3)(1+\omega)(m+k t)^2+2 m\right] }
\end{equation}

\begin{equation}
\label{17}
\Lambda=\xi (m+k t)^2 t^{-2}
\end{equation}  

By using Eq. (\ref{5}), Eq. (\ref{6}) and Eq. (\ref{10}), for $f(R, T) = R+2 \mu T$ model, we obtained  trace  Ricci scalar of dark 
energy matter distribution as:

\begin{equation}
\label{18}
R=6 \left(\frac{m-2 m^2-2k^2 t^2-4m k t}{t^2}\right)
\end{equation}  

\begin{equation}
\label{19}
T^{DF}=(1-3\omega)\left(\frac{(\xi-3)(1+\omega)(m+kt)^2+2m}{\mu (\omega^2-1)t^2}\right)
\end{equation}  
Using Eq. (\ref{18}) and Eq. (\ref{19}), we get $ f(R,T)=R+2\mu T $ as
\begin{equation}
\label{20}
f(R,T)=6 \left(\frac{m-2 m^2-2k^2 t^2-4m k t}{t^2}\right)+2(1-3\omega)\left(\frac{(\xi-3)(1+\omega)(m+kt)^2+2m}{(\omega^2-1)t^2}\right)
\end{equation} 

%%%%%%%%%%%%%%%%%%%%%%%%%%%%%%%%%%%%%%%%%%%%%%%%%%%%%%%%%%%%%%% Figure 1 %%%%%%%%%%%%%%%%%%%%%%%%%%%%%%%%%%%%%%%%%%%%%%%%%%%%%
\begin{figure}[t!]
	\centering
	\includegraphics[scale=0.80]{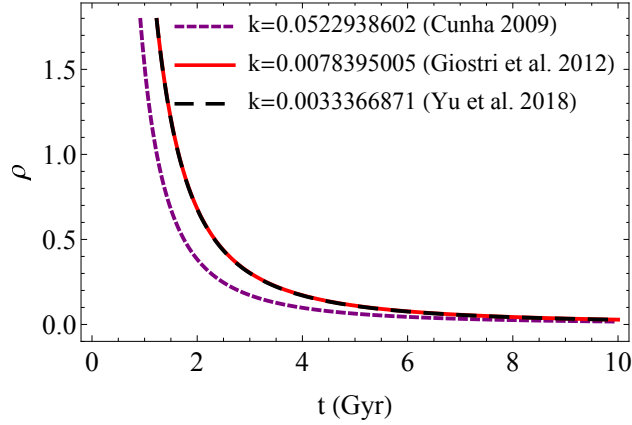}
	\caption{Variation of Energy density $ \rho $ versus time $ t $ with $m=2.0, \mu=-6, \xi=3.75, \omega=-2/3 $.}\label{fig1}	
\end{figure}
%%%%%%%%%%%%%%%%%%%%%%%%%%%%%%%%%%%%%%%%%%%%%%%%%%%%%%%%%%%%%%%%%%%%%%%%%%%%%%%%%%%%%%%%%%%%%%%%%%%%%%%%%%%%%%%%%%%%%%%%%%%%%%%%%
%%%%%%%%%%%%%%%%%%%%%%%%%%%%%%%%%%%%%%%%%%%%%% Figure 2 %%%%%%%%%%%%%%%%%%%%%%%%%%%%%%%%%%%%%%%%%%%%%%%%%%%%%%%%%%%%%%%%%%%%%%
\begin{figure}[t!]
	\centering
	\includegraphics[scale=0.80]{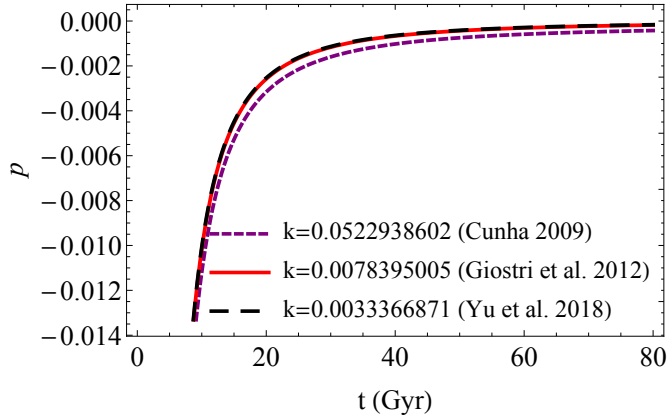}
	\caption{Variation of pressure $ p $ versus time $ t $ with $ m=2.0, \mu=-6, \xi=3.75, \omega=-2/3 $. }\label{fig2}	
\end{figure}
%%%%%%%%%%%%%%%%%%%%%%%%%%%%%%%%%%%%%%%%%%%%%%%%%%%%%%%%%%%%%%%%%%%%%%%%%%%%%%%%%%%%%%%%%%%%%%%%%%%%%%%%%%%%%%%%%%%
%%%%%%%%%%%%%%%%%%%%%%%%%%%%%%%%%%%%%%%%%%%%%%%%%%% Figure 3 %%%%%%%%%%%%%%%%%%%%%%%%%%%%%%%%%%%%%%%%%%%%%%%%%%%%%%%%%
\begin{figure}[t!]
	\centering
	\includegraphics[scale=0.80]{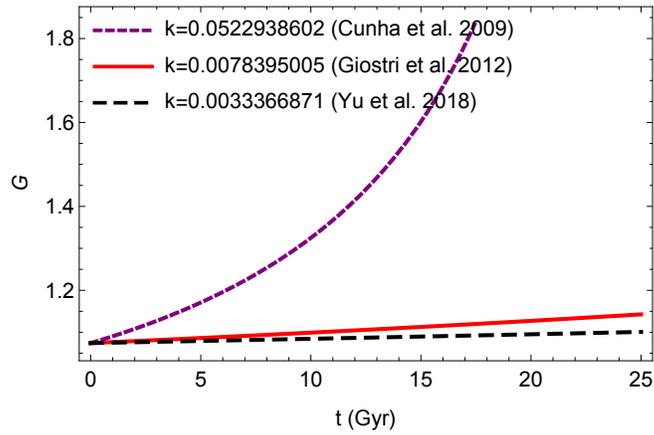}
	\caption{Variation of gravitational term $ G $ versus time $ t $ with $m=2.0, \mu=-6, \xi=3.75, \omega=-2/3 $.}\label{fig3}	
\end{figure}
%%%%%%%%%%%%%%%%%%%%%%%%%%%%%%%%%%%%%%%%%%%%%%%%%%%%%%%%%%%%%%%%%%%%%%%%%%%%%%%%%%%%%%%%%%%%%%%%%%%%%%%%%%%%%%%%%%%%%%%%%%
%%%%%%%%%%%%%%%%%%%%%%%%%%%%%%%%%%%%%%%%%%%%%%%%%%%%%%%%% Figure 4 %%%%%%%%%%%%%%%%%%%%%%%%%%%%%%%%%%%%%%%%%%%%%%%%%%%%%%
\begin{figure}[t!]
	\centering
	\includegraphics[scale=0.80]{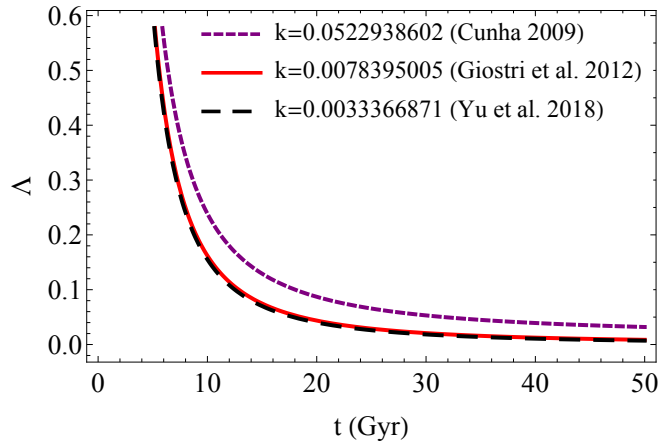}
	\caption{Variation of cosmological term $ \Lambda $ versus time $ t $ with $m=2.0, \mu=-6, \xi=3.75, \omega=-2/3 $.}\label{fig4}	
\end{figure}
%%%%%%%%%%%%%%%%%%%%%%%%%%%%%%%%%%%%%%%%%%%%%%%%%%%%%%%%%%%%%%%%%%%%%%%%%%%%%%%%%%%%%%%%%%%%%%%%%%%%%%%%%%%%%%%%%%%%%%%%%%%%%%%%%%%%%%

Figure $1$, referring to Eq. (\ref{14}) delineates the variety of the energy density $\rho$ versus cosmic time $t$ for three observational data. 
From the figure, we see that $\rho$ is a positive diminishing function of time $t$ and it, converges to zero as $t\to\infty$. Our obtained 
results harmonize with the most recent observations. It is worth mentioning that $\rho$ in case I (Cunha  \cite{ref124}) decreases quickly in comparison 
with cases II (Giostee et al. \cite{ref125}) and case III (Yu at al. \cite{ref127}). \\
 
Figure $2$ corresponding to Eq. (\ref{15}) exhibits a variety of fluid pressure $p$ versus cosmic time $t$ for all three observations. 
Here we found that the isotropic pressure is negative throughout the development. From the figure, we have seen that pressure is a 
decreasing function of cosmic time $t$ and it tends to zero at an early stage. The graph shows that pressure $p$ is high at the beginning 
time and it decreases as time will increase. This negative pressure actually causes the accelerated expansion of the universe for all three 
recent observations.\\

Figure $3$, corresponding to Eq. (\ref{16}) depicts the gravitational term $G(t)$ is zero initially and increases slowly, and leads to infinity.
In the late time $m > 0$, it is easy to see that the gravitational term increases with time, for all three observations. These results are 
consistent with observations. The graph shows that gravitational term $G$ increases adversely with the expansion of cosmic time $t$ and gets 
zero for some late  epochs in all instances.\\

Here in this model the cosmological constant $\Lambda$ determines the nature of the universe. Fig. $4$ explain the cosmological term $\Lambda$ 
vs. cosmic time  $t$. We see that in the present epoch $\Lambda$ is a decreasing function of time $t$, and it has acquired a small positive value. 
Recent cosmological findings \cite{ref128}-\cite{ref129} indicate that our universe could accelerating one. Thus, the nature of  $\Lambda$ in our 
derived models is supported by observations. However, as compared to radiating dominated and rigid fluid, it decreases more sharply with cosmic 
time an empty universe. In the radiating dominated universe,  $\Lambda$ term often decreases.\\

 %%%%%%%%%%%%%%%%%%%%%%%%%%%%%%%%%%%%%%%%%%%%%%%%%%%%%%%%%%%%%%Section 4 %%%%%%%%%%%%%%%%%%%%%%%%%%%%%%%%%%%%%%%%%%%%%%%%%%%%%%%%%%%
 
\section{ Correspondence with scalar field models in f(R, T) gravity}
The DE models in the scalar field are already part of the family of dynamic of dark energy models that describe the phenomenon of DE.
The literature contains a wide variety of such models as tachyon, $k$-essence, quintessence, phantom, ghost dilation, and condensate, etc.\\

Reconstituting the capacity models of DE would be necessary to clarify the cosmological nature of quantum gravity via scalar fields. 
Such models depict the universe's quintessence behavior and also provide an effective overview of DE. In this section, we explore the 
field of tachyon, k-essence and, quintessence DE models in $f(R, T)$ theory with variable $G$ and $\Lambda$.  Here we consider the value of 
EoS parameter is $ \omega=-1/3$ and $-2/3$ with reference to redshift. Now we plot kinetic energy  and scalar potential with respect to $z$ 
by using $a= a_{0}(1+z)^{-1}$ in all cases, here we assume $ a_{0}=1$ used many authors \cite{ref130,ref131}

%%%%%%%%%%%%%%%%%%%%%%%%%%%%%%%%%%%%%%%%%%%%%%%%%%%%%%%%%%%%%%% Subsection 4.1 %%%%%%%%%%%%%%%%%%%%%%%%%%%%%%%%%%%%%%%%%%%%%%%%%%%%
\subsection{Tachyon field with redshift}

The tachyon field was suggested to be the source of dark energy\cite{ref69,ref87} and may be described by effective field theory corresponding
to some kind of tachyon condensate with an effective Lagrangian density given by \cite{ref88,ref67}. This field is one of the DE components 
which describes the accelerated expansion of the universe. The EoS parameter of the tachyon DE matter distribution lies between $-1$ and $0$ 
Gibbon \cite{ref66}. The energy density $ \rho $ and  pressure $ p $ in  flat FRW background for tachyon matter distribution related to SF 
($ \Phi $) and scalar potential $ V(\Phi) $ are given as:

\begin{equation}
\label{21}
p_{TF}=-T^{i}_{i}=V(\Phi)\sqrt{1-\dot{\Phi}^2}
\end{equation}

{\begin{equation}
\label{22}
\rho_{TF}=T^{4}_{4}=V(\Phi) \left(1-\dot{\Phi}^2\right)^{-1/2}
\end{equation}

Here, $ \dot{\Phi}^2 $ is the kinetic energy (KE) and $ V(\Phi) $ is the scalar potential for a given scalar field. In redshift cut-off 
$\Phi$ and $V(\Phi)$ obtained as:

%%%%%%%%%%%%%%%%%%%%%%%%%%%%%%%%%%%%%%%%%%%%%%%%%%%%%%%%%% Figure 5 %%%%%%%%%%%%%%%%%%%%%%%%%%%%%%%%%%%%%%%%%%%%%%%%%%%%%%%%%%

\begin{figure}[t!]
	\centering
	\includegraphics[scale=0.60]{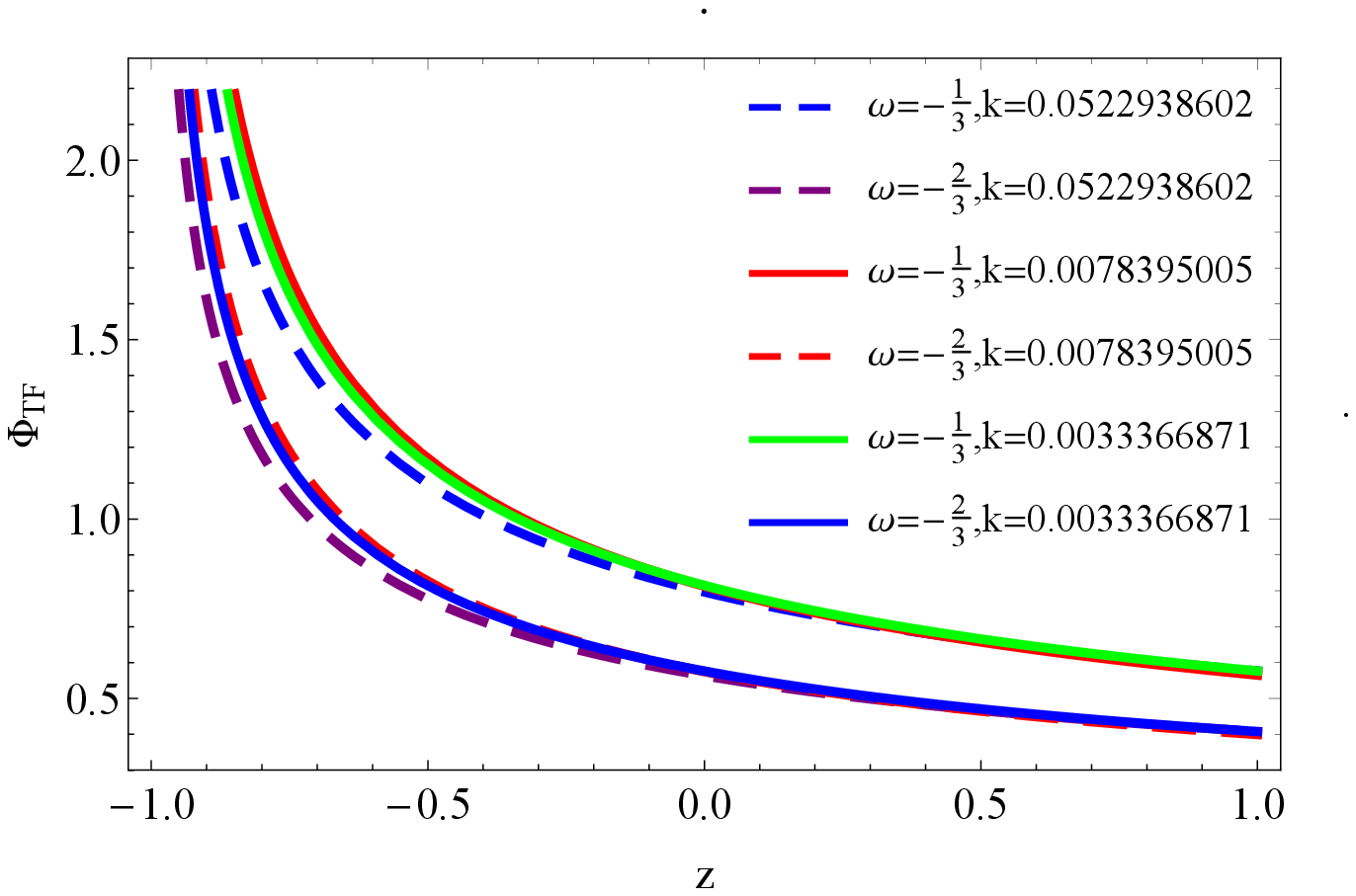}
	\caption{Variation of SF for tachyon versus Redshift $ z $  with $m=2, \mu=-6, \xi=3.75 $ and $ c_{1}=0 $. }\label{fig5}	
\end{figure}
%%%%%%%%%%%%%%%%%%%%%%%%%%%%%%%%%%%%%%%%%%%%%%%%%%%%%%%%%%%%%%%%%%%%%%%%%%%%%%%%%%%%%%%%%%%%%%%%%%%%%%%%%%%%%%%%%%%%%%%%%%%%%%%%%%%%%%%%
%%%%%%%%%%%%%%%%%%%%%%%%%%%%%%%%%%%%%%%%%%%%%%%%%%%%% Figure 6 %%%%%%%%%%%%%%%%%%%%%%%%%%%%%%%%%%%%%%%%%%%%%%%%%%%%%%%%%%%%%%%%%%%%%%%%

\begin{figure*}[t!]
	\centering
	\begin{subfigure}[t]{0.45\linewidth}
		\centering
		\includegraphics[width=\linewidth]{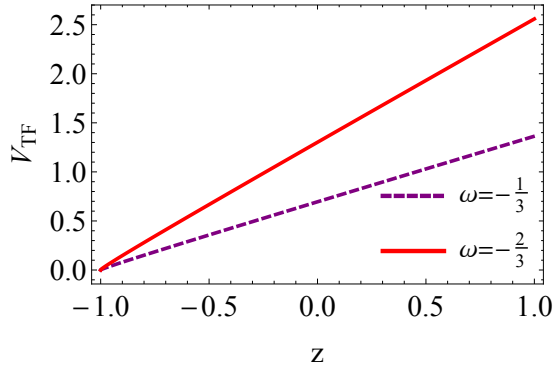}
		\caption{$   k=0.0522938602 $, Cunha  (2009)} 
	\end{subfigure} 
	\quad\quad\begin{subfigure}[t]{0.45\linewidth}
		\centering
		\includegraphics[width=\linewidth]{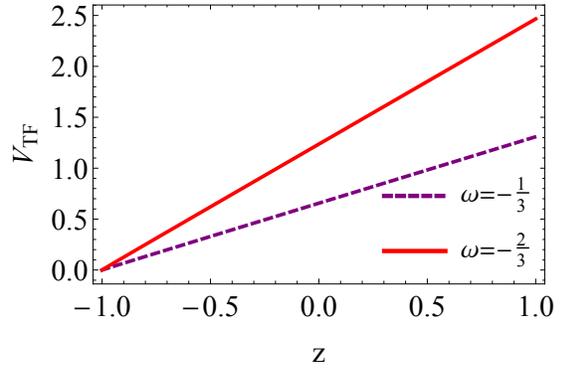}
		\caption{$ k=0.0078395005 $, Giostri et al.(2012)}
	\end{subfigure}\\
	\quad\quad\begin{subfigure}[t]{0.45\linewidth}
		\centering
		\includegraphics[width=\linewidth]{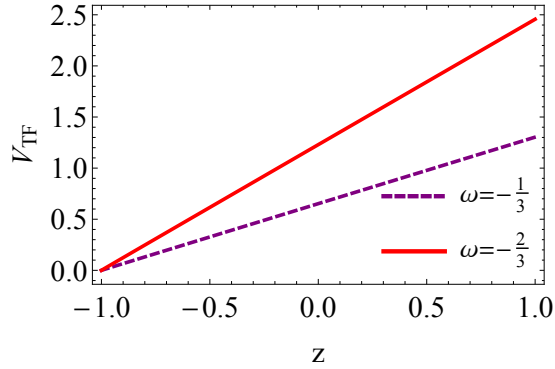}
		\caption{$ k=0.0033366871 $, Yu et al.(OHD+JLA Data 2018)}
	\end{subfigure}
	\caption{Variation of scalar potential for tachyon field versus Redshift $ z $  with $ m=2, \mu=-6, \xi=3.75 $.}\label{fig6}
\end{figure*}
%%%%%%%%%%%%%%%%%%%%%%%%%%%%%%%%%%%%%%%%%%%%%%%%%%%%%%%%%%%%%%%%%%%%%%%%%%%%%%%%%%%%%%%%%%%%%%%%%%%%%%%%%%%%%%%%%%%%%%%%%%%%%%%%%%%%%%%%
\begin{equation}
\label{23}
\Phi_{TF}=\frac{m \sqrt{\omega+1} W\left(\frac{k \left(\frac{1}{z+1}\right)^{1/m}}{m}\right)}{k} +c_{1}
\end{equation}  

\begin{equation}
\label{24}
V(z)_{TF}=\frac{k^2 \sqrt{-\omega } \left((\xi -3) (\omega +1) \left(m W\left(\frac{k \left(\frac{1}{z+1}\right)^{1/m}}{m}\right)+
m\right)^2+2 m\right)}{\mu \ m^2 \left(\omega ^2-1\right) W\left(\frac{k \left(\frac{1}{z+1}\right)^{1/m}}{m}\right)^2}
\end{equation}

Figure $5$, shows the kinetic tachyon energy with respect to redshift $z$ for three recent observations \cite{ref124,ref125,ref126} by 
considering the $\omega = -1/3~ $ \& $~ -2/3$.  Graph represents that kinetic energy is decreases as $z$ increases. We find if 
$z\to-1$, $\Phi$ is very high and $z\to1$ $\Phi$ is very low. Similarly Figure 6(a,b,c) represents the scalar field effect with redshift
$(z)$. For all three  observations $V(z)$ increases with redshift for $\xi>3$. Obtain the following expression for the tachyon potential 
in terms of the redshift we found that the tachyon 
field decreases as the universe expands.

%%%%%%%%%%%%%%%%%%%%%%%%%%%%%%%%%%%%%%%%%%%%%%%%%%%%%%%%%%Subsection 4.2 %%%%%%%%%%%%%%%%%%%%%%%%%%%%%%%%%%%%%%%%%%%%%%%%%%%%%%%
\subsection{k-essence field with redshift}
	 
The k-essence scalar field (SF) model is used to describe the universe's late-time acceleration, described by the string theory Born-Infeld 
action \cite{ref134,ref135}. $k$-essence scenarios are well known attractor-like dynamics, and thus ignore the fine-tuning of the initial scalar 
field conditions\cite{ref136}. In a flat FRW context, the values of pressure $p$ and energy density $\rho$ are given as, for the kinetic and 
scalar potential distribution of k-essence \cite{ref137} . 

\begin{equation}
\label{25}
p_{ke}=-T^{i}_{i}=V(\Phi)\left(\frac{\dot{\Phi}^4}{4}-\frac{\dot{\Phi}^2}{2}\right)
\end{equation}  

\begin{equation}
\label{26}
\rho_{ke}=T^{4}_{4}=V(\Phi)\left(\frac{3\dot{\Phi}^4}{4}-\frac{\dot{\Phi}^2}{2}\right)
\end{equation}

In redshift cut- off $\Phi$ and $V(\Phi)$ of the $k$-essence is obtained as:

\begin{equation}
\label{27}
\Phi_{ke}=\frac{m \sqrt{\frac{2 \omega -2}{3 \omega -1}} W\left(\frac{k \left(\frac{1}{z+1}\right)^{1/m}}{m}\right)}{k}+c_{2}
\end{equation}  
\begin{equation}
\label{28}
V(z)_{ke}=\frac{k^2 (3 \omega -1)^2 \left((3-\xi ) (\omega +1) \left(m W\left(\frac{k \left(\frac{1}{z+1}\right)^{1/m}}{m}\right)+
m\right)^2-2 m\right)}{2 \mu \ m^2 (\omega -1) \left(\omega ^2-1\right) W\left(\frac{k \left(\frac{1}{z+1}\right)^{1/m}}{m}\right)^2}
\end{equation}

%%%%%%%%%%%%%%%%%%%%%%%%%%%%%%%%%%%%%%%%%%%%%%%%%%%%% Figure 7 %%%%%%%%%%%%%%%%%%%%%%%%%%%%%%%%%%%%%%%%%%%%%%%%%%%%%%%%%
\begin{figure}[t!]
	\centering
	\includegraphics[scale=0.60]{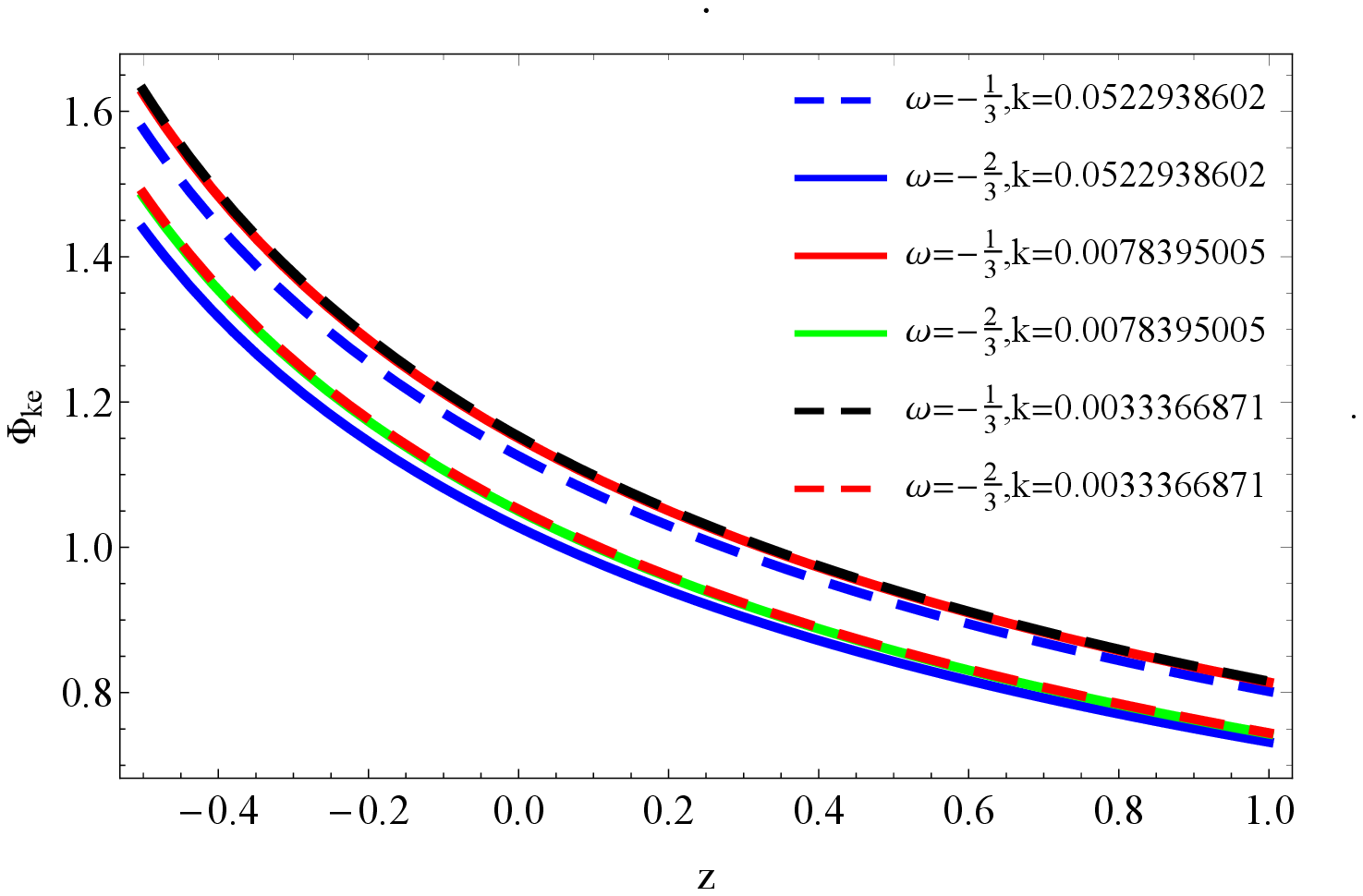}
	\caption{Variation of SF for k-essence versus Redshift $ z $  with $  \mu=-6, \xi=2 $ and $ c_{2}=0 $. }\label{fig7}	
\end{figure}
%%%%%%%%%%%%%%%%%%%%%%%%%%%%%%%%%%%%%%%%%%%%%%%%%%%%%%%%%%%%%%%%%%%%%%%%%%%%%%%%%%%%%%%%%%%%%%%%%%%%%%%%%%%%%%%%%%%%%%%%%%
%%%%%%%%%%%%%%%%%%%%%%%%%%%%%%%%%%%%%%%%%%%%%%%%%%%%%%%%%% Figure 8 %%%%%%%%%%%%%%%%%%%%%%%%%%%%%%%%%%%%%%%%%%%%%%%%%%%%%%%
\begin{figure*}[t!]
	\centering
	\begin{subfigure}[t]{0.45\linewidth}
		\centering
		\includegraphics[width=\linewidth]{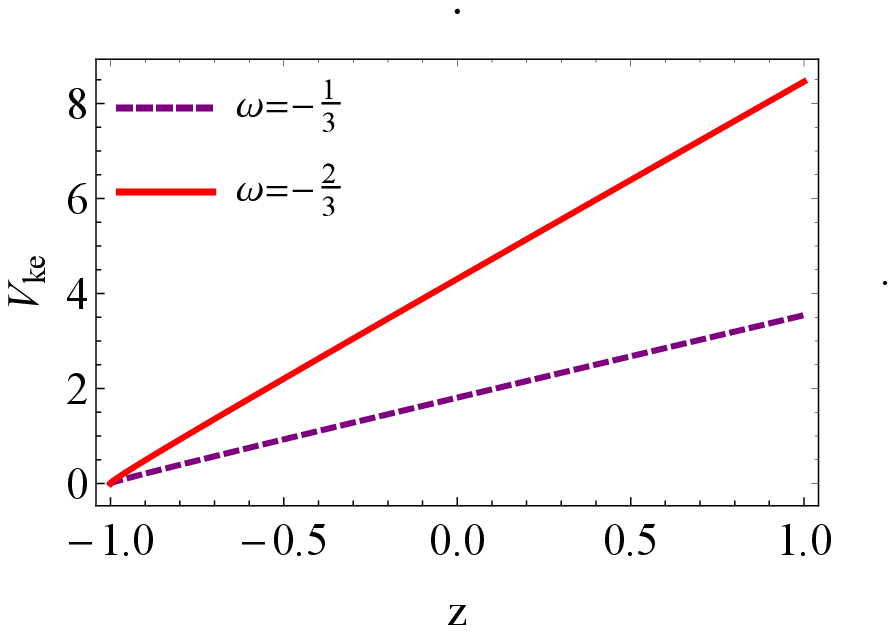}
		\caption{$   k=0.0522938602 $, Cunha  (2009)} 
	\end{subfigure} 
	\quad\quad\begin{subfigure}[t]{0.45\linewidth}
		\centering
		\includegraphics[width=\linewidth]{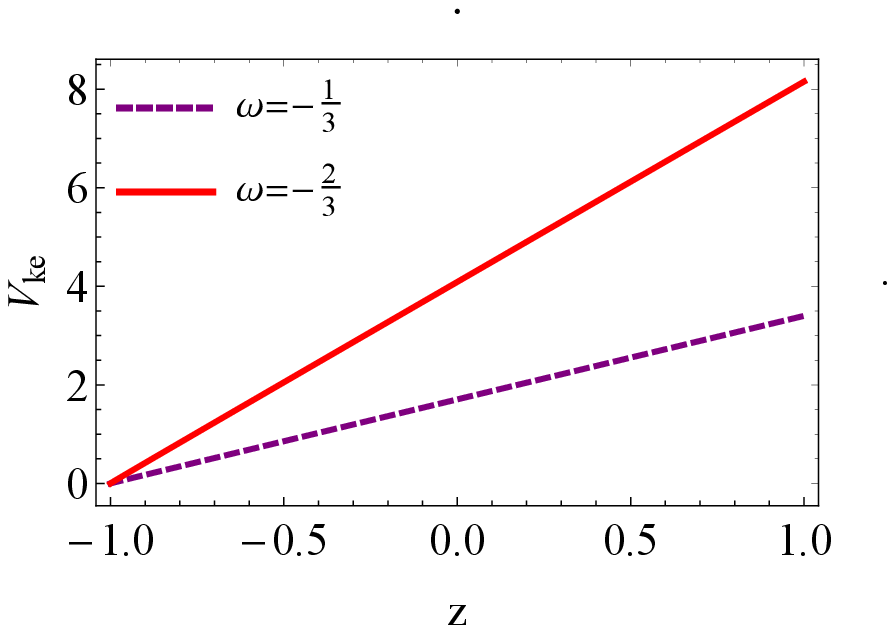}
		\caption{$ k=0.0078395005 $, Giostri et al.(2012)}
	\end{subfigure}\\
	\quad\quad\begin{subfigure}[t]{0.45\linewidth}
		\centering
		\includegraphics[width=\linewidth]{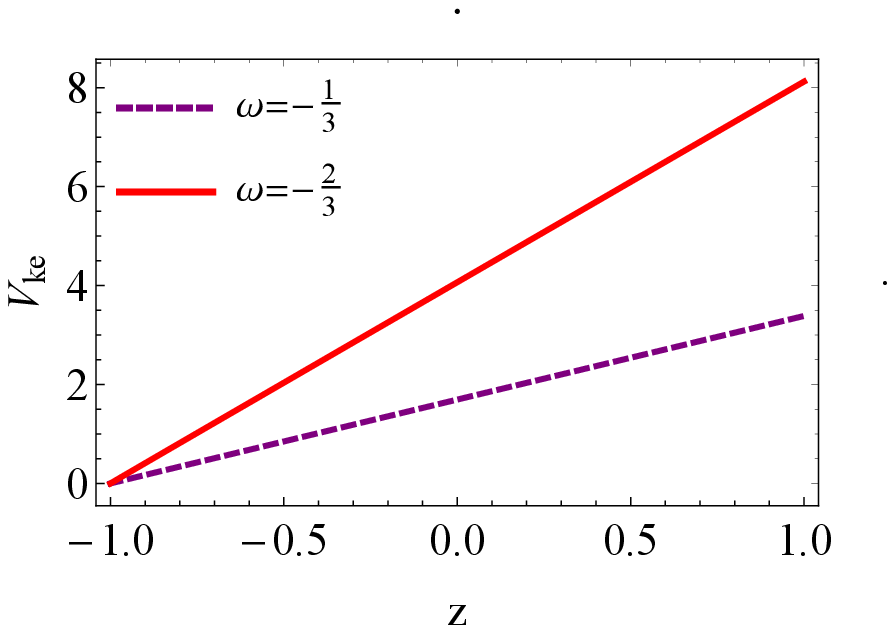}
		\caption{$ k=0.0033366871 $, Yu et al.(OHD+JLA Data 2018)}
	\end{subfigure}
	\caption{Variation of scalar potential for k-essence versus Redshift $ z $  with $ m=2, \mu=-6, \xi=3.75 $.}\label{fig8}
\end{figure*}
%%%%%%%%%%%%%%%%%%%%%%%%%%%%%%%%%%%%%%%%%%%%%%%%%%%%%%%%%%%%%%%%%%%%%%%%%%%%%%%%%%%%%%%%%%%%%%%%%%%%%%%%%%%%%%%%%%%%%%%%%%%%%%%%%%%

Figure $7$, shows the kinetic k-essence energy with respect to redshift $z$ for three recent observations \cite{ref124,ref125,ref126} by 
considering the $\omega = -1/3 ~ $\& $ ~-2/3$. The behavior of k-essence field is similar to tachyon field. Our graph represents the 
kinetic energy is decreases as $z$ increases. Similarly Figure 8(a,b,c). represents the scalar potential effect with redshift $(z)$. For 
all three  observations it increases with redshift for $\xi>3$.

%%%%%%%%%%%%%%%%%%%%%%%%%%%%%%%%%%%%%%%%%%%%%%%%%%%%%%%%%%%%% Subsection 4.3 %%%%%%%%%%%%%%%%%%%%%%%%%%%%%%%%%%%%%%%%%%%
\subsection{Quintessence field with redshift}

%%%%%%%%%%%%%%%%%%%%%%%%%%%%%%%%%%%%%%%%%%%%%%%% Figure 9 %%%%%%%%%%%%%%%%%%%%%%%%%%%%%%%%%%%%%%%%%%%%

\begin{figure}[t!]
	\centering
	\includegraphics[scale=0.60]{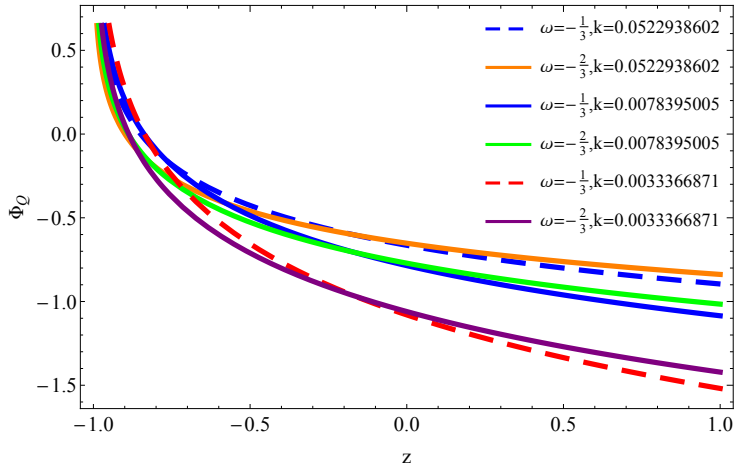}
	\caption{Variation of SF for Quintessence versus Redshift $ z $ with $ m=2, \mu=-6, \xi=3.75 $ and $ c_{3}=0 $. }\label{fig9}	
\end{figure}
%%%%%%%%%%%%%%%%%%%%%%%%%%%%%%%%%%%%%%%%%%%%%%%%%%%%%%%%%%%%%%%%%%%%%%%%%%%%%%%%%%%%%%%%%%%%%%%%%%%%%%%%%%%%%%%%%%%%%%%%

Quintessence is a hypothetical form of dark energy, which is described by homogeneous and time-dependent SF, as well as the scalar potential 
which leads to universe acceleration \cite{ref135}. EoS quintessence parameter indicates the accelerated universe expansion within the  
$-1\leq\omega \leq -\frac{1}{3}$ \cite{ref135}. In flat FRW universe, the relations of energy density  and the 
pressure respectively, in terms of quintessence SF and scalar potential is given as \cite{ref107}:

\begin{equation}
\label{29}
p_{Q}=-T^{i}_{i}=\frac{\dot{\Phi}^2}{2}-V(\Phi)
\end{equation}  

\begin{equation}
\label{30}
\rho_{Q}=T^{4}_{4}=\frac{\dot{\Phi}^2}{2}+V(\Phi)
\end{equation}
%%%%%%%%%%%%%%%%%%%%%%%%%%%%%%%%%%%%%%%%%%%%%%%%%%%%%%%%%%%%% Figure 10 %%%%%%%%%%%%%%%%%%%%%%%%%%%%%%%%%%%%%%%%%%%%%%%

\begin{figure*}[t!]
	\centering
	\begin{subfigure}[t]{0.45\linewidth}
		\centering
		\includegraphics[width=\linewidth]{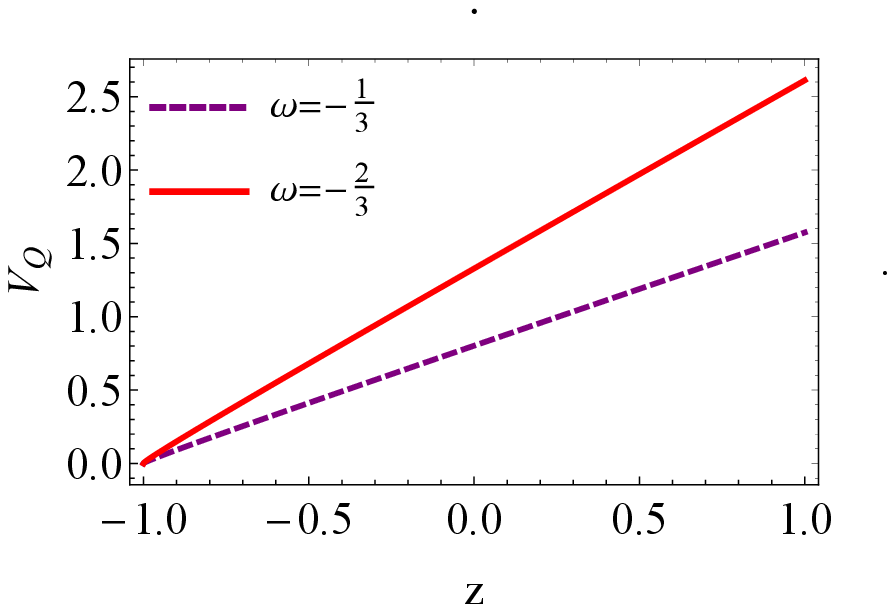}
		\caption{$   k=0.0522938602 $, Cunha  (2009)} 
	\end{subfigure} 
	\quad\quad\begin{subfigure}[t]{0.45\linewidth}
		\centering
		\includegraphics[width=\linewidth]{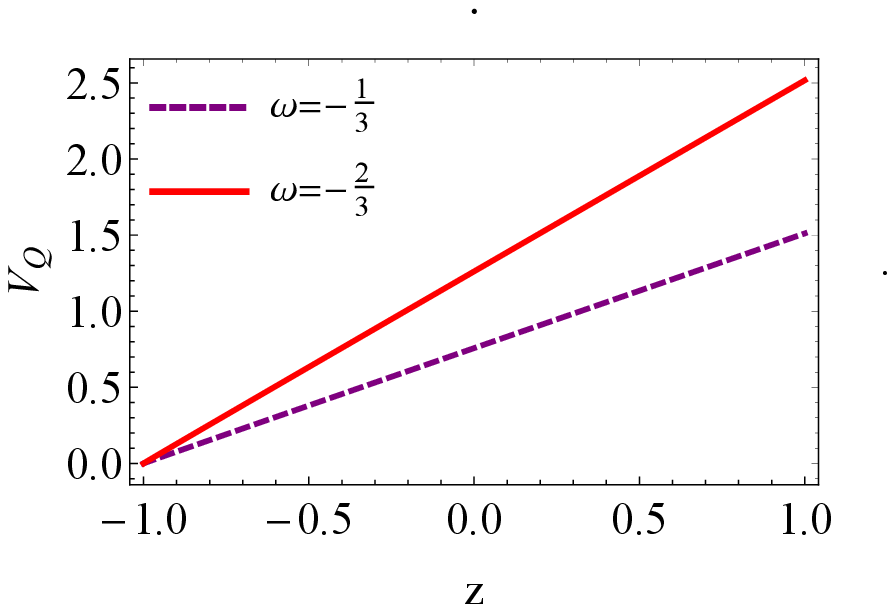}
		\caption{$ k=0.0078395005 $, Giostri et al.(2012)}
	\end{subfigure}\\
	\quad\quad\begin{subfigure}[t]{0.45\linewidth}
		\centering
		\includegraphics[width=\linewidth]{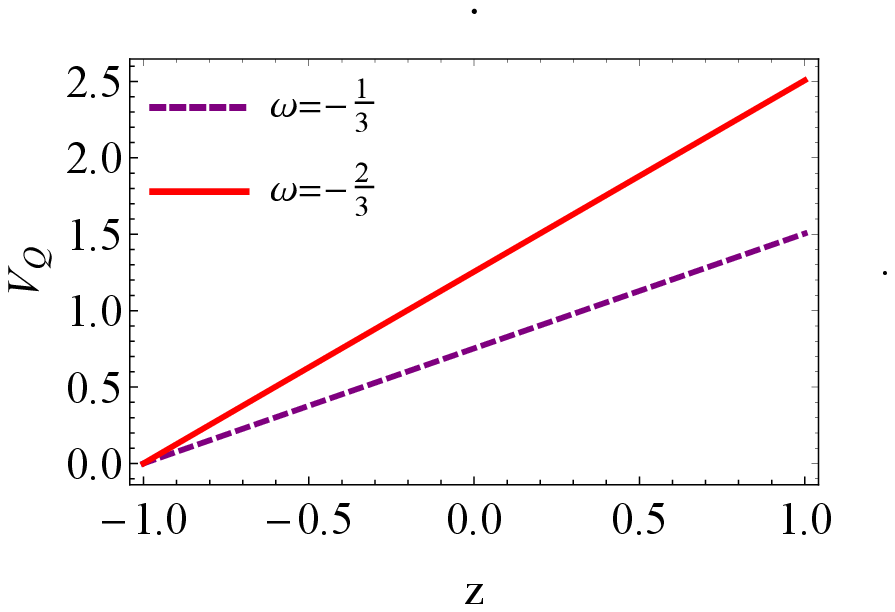}
		\caption{$ k=0.0033366871 $, Yu et al.(OHD+JLA Data 2018)}
	\end{subfigure}
	\caption{Variation of scalar potential for quintessence versus $ z $  with $ m=2, \mu=-6, \xi=3.75 $.}\label{fig10}
\end{figure*}
%%%%%%%%%%%%%%%%%%%%%%%%%%%%%%%%%%%%%%%%%%%%%%%%%%%%%%%%%%%%%%%%%%%%%%%%%%%%%%%%%%%%%%%%%%%%%%%%%%%%%%%%%%%%%%%%%%%%%%%%%%%
In redshift cut- off $\Phi$ and $V(\Phi)$ of the quintessence is obtained as:
\begin{equation}
\label{31}
\dot{\Phi_{Q}}=\sqrt{\frac{k^2 \left((\xi -3) (\omega+1) \left(m W\left(\frac{k \left(\frac{1}{z+1}\right)^{1/m}}{m}\right)+
m\right)^2+2 m\right)}{\mu  m^2 (\omega-1) W\left(\frac{k \left(\frac{1}{z+1}\right)^{1/m}}{m}\right)^2}}
\end{equation}  
On integration Eq. (\ref{31}), we get
\begin{eqnarray}
\label{32}
&\Phi_{Q}=\frac{a_{3} \log (m a_{1})+m\log\left(m a^{2}_{4} (a_{1}+1)\right)+a_{2} a_{4}}{\sqrt{\mu (\omega-1)}}+ \nonumber\\
&\frac{a_{2}-a_{3}\log \left(m^2 a^{2}_{4}+a_{1} a_{3}+a_{3} \sqrt{m^2 a^{2}_{4}(1+a_{1})^2+2 m }\right)}{\sqrt{\mu (\omega-1)}}+c_{3},
\end{eqnarray}  
where, 
\[ a_{1}=W\left(\frac{k \left(\frac{1}{z+1}\right)^{1/m}}{m}\right) \] 

\[ a_{2}=\sqrt{k^2 \left((\xi -3) (\omega+1) \left(m W\left(\frac{k \left(\frac{1}{z+1}\right)^{1/m}}{m}\right)+m\right)^2+2 m\right)} \]
\[a_{3}=\sqrt{2 m+m^2 (1+\omega)(\xi-3)}\] 
\[a_{4}=\sqrt{(\omega+1)(\xi-3)}\]

\begin{equation}
\label{33}
V(z)_{Q}=\frac{k^2 \left((3-\xi ) (\omega +1) \left(m W\left(\frac{k \left(\frac{1}{z+1}\right)^{1/m}}{m}\right)+
m\right)^2-2 m\right)}{2 \mu  m^2 (\omega +1) W\left(\frac{k \left(\frac{1}{z+1}\right)^{1/m}}{m}\right)^2}
\end{equation}

Figure 9, shows the kinetic quintessence energy with respect to redshift $z$ for three recent observations \cite{ref78,ref79,ref80} by 
considering the $\omega = -1/3 ~  $\&$ ~-2/3$. The behaviour of the quintessence field is similar to k-essence and tachyon field. Our 
graph represents the kinetic energy is decreases as $z$ increases. Similarly Figure 10(a,b,c) represents the Scalar potential effect 
with redshift $(z)$. For all three  observations it increases with redshift for $\xi>3$.

%%%%%%%%%%%%%%%%%%%%%%%%%%%%%%%%%%%%%%%%%%%%%%%%%%%%%%%%%%%%% Section 5 %%%%%%%%%%%%%%%%%%%%%%%%%%%%%%%%%%%%%%%%%%%%%%%%%%

\section{Conclusion}
In the present manuscript, the scalar fields are playing a very important role in cosmological consequence. Particularly cosmological 
 inflation, the late-time acceleration of the universe, or dark matter, and its properties can be clarified in the sense of unique scalar 
 field models. In this model, we considered a generalized gravity model with an arbitrary coupling between matter (described by the trace of 
the stress-energy tensor) and geometry, with the Lagrangian given by $T$ and the Ricci scalar arbitrary function.
Here we derived the corresponding gravitational field equations and analyzed some particular cases that may be important to understand 
some open problems of cosmology and astrophysics'. In this paper, we have investigated various DE candidates in $f(R, T)$ gravitation theory 
for flat FRW universe with varying $G$ and $\Lambda$ by using the latest findings.\\

The main outcomes of the derived  models are  summarized as: \\
\begin{itemize} 
\item In our model Fig. $1$ depicts the energy density $\rho$ is infinitely large at the initial stage and tends to zero as $t\to\infty$, which is 
 consistent with the Big Bang theory. 
 
 \item Fig. $2$ illustrates that pressure $p$ is a decreasing function of time. This negative pressure actually causes the accelerated 
 expansion of the universe for all the three recent observations. 
 Also $t\to\infty$, $\xi=3$, we get $G\to\frac{\mu(\omega-3)}{4\pi}$  in $f(R, T)$ theory.
 
 \item Gravitational term is initially zero and it increases as $t$ increases (Fig. 3). Different kinds of tests, experiments, and measurements 
 were involved to verify the variation of the gravitational constant with time. 
 
 \item Schmidt et al. \cite{ref94} propose that 
 $\Lambda$ with a magnitude of $\Lambda(Gh / c^{3})=10^{-23}$ has positive cosmological constants. Such findings on the magnitude and 
 the redshift of type Ia supernova indicates that through the cosmological $\Lambda-term$ our universe can be accelerating with induced 
 cosmological density. 
 
 \item Finally, we studied the relationship between scalar field DE models in Figs. $5$-$10$ involving quintessence, tachyon, and k-essence. 
 Here we have subsequently explored the scalar fields and given particular attention to dark energy models by using a constant EoS 
 parameter $\omega= -1/3, -2/3$
 
\item Figs. $9$-$10$ displays the evolutionary trajectories of the $\Phi$ kinetic energy and the related scalar potential concerning the redshift $z$.
We notice that the quintessence field increases with a high redshift. Similarly, the evolution of tachyon and k-essence fields is very much 
similar to quintessence. 

\item We may conclude that our model begins with Big Bang and terminate with Big Rip. This outcome is a consequence of the significant shift 
in the evolution of the transition between matter and dark energy dominated epochs. The positive values of scalar potential are obtained for 
negative values of $\mu$. In addition, for $\mu = 0, $, we get $G = 0$ GRT solutions with FRW universe the tachyon field, k-essence, and 
quintessence matter distributions are obtained. Such DE solutions minimize the importance of GRT. We wish to point out that now the results 
are the current state of the universe is consistent with the scalar field and corresponding potential. We have formulated the potential and the
dynamics of these scalar field models representing tachyon, k-essence, and quintessence cosmology.

\end{itemize}
 
Hence, the solutions exhibited in this work can be helpful for better comprehension of the features of FRW models
in the Universe evolution in $f(R, T)$ gravity for tachyon field, k-essence field, and quintessence field with redshift.

%%%%%%%%%%%%%%%%%%%%%%%%%%%%%%%%%%%%%%%%%%%%%%%%%%%%%%%%%%%%%%%%%%%%%%%%%%%%%%%%%%%%%%%%%%%%%%%%%%%%%%5
\section*{Acknowledgments} 
The authors would like to thank Director Prof. Anoop K. Gupta for providing facility and support at 
GLA University. The authors also thank Dr. Shiva Durga for reading the manuscript and giving her valuable 
suggestions.

%%%%%%%%%%%%%%%%%%%%%%%%%%%%%%%%%%%%%%%%%%%%%%%%%%%%%%%%%%%%%%%%%%%%%%%%%%%%%%%%%%%%%%%%%%%%%%%%%%%%%%5

%%%%%%%%%%%%%%%%%%%%%%%%%%%%%%%%%%%%%%%%%%%%%%%%%%%%%%%%%%%%%%%%%%%%%%%%%%%%%%%%%%%%%%%%%%%%%%%%%%%%%%%%%%%%%%%%%%%%%%%%%%%%%%5 

\end{document}